\documentclass[11pt]{article}

\usepackage[utf8]{inputenc}
\usepackage[T1]{fontenc}
\usepackage[english]{babel}

\usepackage{indentfirst}

\usepackage{amsmath, amssymb, amsfonts}

\usepackage{graphicx}
\usepackage{float}

\usepackage{authblk}

\usepackage{geometry}
\usepackage{hyperref}
\hypersetup{
    colorlinks=true,
    linkcolor=blue,
    citecolor=blue,
    urlcolor=blue
}

\usepackage[backend=biber,style=numeric,sorting=none,doi=true,url=false,isbn=false]{biblatex}
\DeclareFieldFormat{doi}{%
  \href{https://doi.org/#1}{doi:#1}%
}

\title{A note on decoherence in quantum cosmology by thermal perturbations: an open quantum systems approach}

\author[a]{Alexander Kulyabin \thanks{alexander.kulyabin@uni-bayreuth.de}}

\affil[a]{ {\small \textit{Theoretical Physics VI, University of Bayreuth, Bayreuth, Germany }}}

\date{}

\begin{document}

\maketitle

\begin{abstract}
We study, within a toy model, how a thermal bath of cosmological perturbations can decohere the minisuperspace degrees of freedom of the universe, treating the perturbations as an environment in the framework of open quantum systems. Instead of tracing out the perturbations from a semiclassical ansatz for the total wave function, as in earlier approaches, we incorporate their backreaction on the background through an effective second-order coupling and derive a master equation. The model describes a spatially flat semiclassical universe during slow-roll inflation with a dust clock field, while scalar cosmological perturbations in a phenomenologically assumed thermal state constitute the environment. The master equation is non-Markovian, since the bath correlation functions do not decay in time. Within this framework, we find that, for a representative set of parameter estimates, the characteristic decoherence timescale corresponding to the thermal contribution can be shorter than the Hubble timescale. In addition, we address the question of whether the pointer-state structure induced by the decoherence in such a framework could provide a criterion for the choice of semiclassical states in quantum cosmology.
\end{abstract}

\section{Introduction}
\label{1}
One of the most fundamental and conceptual questions of quantum cosmology is the quantum-to-classical transition of the universe
\cite{10.1093/acprof:oso/9780199212521.001.0001}. In particular, an important objective is to understand how the classical observable regime may emerge during inflation. The idea that the mechanism of decoherence could provide such a transition goes back to the pioneering works
\cite{PhysRevD.31.1777, Kiefer_1987, Halliwell:1989vw, Barvinsky_1998}. Within this framework, the homogeneous cosmological degrees of freedom become entangled with inhomogeneous perturbations, and tracing over the latter suppresses quantum interference in the reduced density matrix of the cosmological sector.

These ideas established one of the standard approaches to decoherence in quantum cosmology
\cite{novikov2021quantumdecoherencehomogeneousmodes, Barvinsky_2022}. A key assumption is that the total wave function can be approximated by a product of a prescribed wave function of the homogeneous universe (for example, the Hartle-Hawking or Vilenkin state) and a wave function describing inhomogeneous perturbations within the semiclassical approximation. The perturbative degrees of freedom are then traced out, yielding a reduced density matrix for the homogeneous sector.

In the present work, we explore an alternative description based on the formalism of open quantum systems
\cite{10.1093/acprof:oso/9780199213900.001.0001}. In this picture, decoherence may arise from an effective interaction between the cosmological degrees of freedom and perturbations regarded as an environment. Coherence may be lost through the interaction with a large number of environmental degrees of freedom, leading to a suppression of interference in the reduced state of the system. During the comparatively short decoherence timescale considered here, the perturbations are assumed to remain in a sufficiently stable state. This formulation allows one to employ the standard language of master equations
\cite{10.1093/acprof:oso/9780199213900.001.0001, Manzano_2020}, together with the associated concepts and techniques developed in the theory of open quantum systems.

An interesting question in this context concerns the relation between decoherence and the choice of the semiclassical state of the universe. This issue is particularly relevant in effective models of loop quantum cosmology, where coherent states are commonly adopted when evaluating quantum corrections of the dynamics
\cite{Taveras_2008} and the comparison of different classes of semiclassical states has itself been a subject of investigation \cite{Corichi_2011}. It is conceivable that the quantum evolution itself could provide a criterion for selecting preferred semiclassical states. Within the language of open quantum systems, this idea is naturally related to the notion of pointer states, namely those states that remain most robust under decoherence
\cite{zurek2003decoherencetransitionquantumclassical, Zurek_2022}.

Open quantum system formalism has been extensively considered in the context of cosmological perturbations themselves, treated as an open system on a fixed classical
background \cite{burgess2014efthorizonstochasticinflation, Martin_2018, Colas_2022, Burgess_2023}.  Moreover, the formalism of open quantum systems has previously been applied to models of gravitationally induced decoherence
\cite{Anastopoulos_2013, Blencowe_2013}, where thermal baths of metric perturbations interacting with quantum fields were considered. More recently, related ideas have also been explored in the contexts of neutrino oscillations and effective quantum-mechanical models
\cite{Domi_2024, fahn2026generalisinggravitationallyinduceddecoherence}. In the present work the perturbations constitute the environment, while the system is the
homogeneous gravitational background itself.

An essential assumption of the present toy model is that the environment is described by a thermal state. Our motivation for such a consideration comes from studies of thermal generalizations of the Bunch-Davies vacuum for inflationary perturbations
\cite{li2019preinflationaryperturbationsthermalbackground, Bhattacharya_2006, Wang_2017}. Such scenarios have been argued to be compatible with current observations for sufficiently low effective temperatures and have been suggested as a possible explanation of the large-scale power deficit observed in the CMB spectrum
\cite{li2019preinflationaryperturbationsthermalbackground}. We emphasize that this assumption is introduced phenomenologically and is not derived within the present work. Whether this assumption provides the most appropriate description of the perturbation state remains an open question and deserves further investigation.

For the interaction between the cosmological sector and the perturbations, we consider the backreaction formalism
\cite{Mukhanov_1997, li2008cosmological}. Within this approach, the influence of perturbations on the homogeneous background is described through an effective energy-momentum tensor obtained from the effective averaging of the second-order expansion of the Einstein equations. The perturbation dynamics themselves remain unchanged to the considered order, making this framework compatible with the assumption that the influence of the system on the environment is relatively weak. We quantize the corresponding backreaction contribution using operators of the cosmological system together with creation and annihilation operators of the perturbations and subsequently trace over the environmental degrees of freedom to derive the master equation.

A further conceptual issue in applying the theory of open quantum systems to quantum cosmology is the well-known problem of time in canonical quantum gravity
\cite{10.1093/acprof:oso/9780199212521.001.0001, anderson2012problemtimequantumgravity}. To overcome this issue, we consider an additional dust clock field
\cite{Brown_1995, Husain_2012}, which provides one of the technically simplest realizations of a physical time variable suitable for the present analysis.

We construct a toy model describing a spatially flat universe with a scalar inflaton during the slow-roll exponential expansion in the presence of such a clock field. Following
\cite{Ali_2018}, the minisuperspace dynamics reduce to those of a harmonic oscillator with an inverted potential. For the environment, we consider scalar cosmological perturbations in the super-horizon regime, corresponding to the assumed low effective temperature. In this limit, the backreaction of the perturbations is described by an effective cosmological constant
\cite{brandenberger2002reactioncosmologicalperturbationscosmological}.

Our primary objective is to derive the corresponding master equation and to investigate the qualitative properties of the resulting decoherence mechanism. We then present representative order-of-magnitude estimates for the thermal decoherence rate. These estimates should be regarded as illustrative rather than definitive, since they depend on the assumptions underlying the toy model. A more systematic investigation of the dependence on the cutoffs may be worthwhile and is left for future work.

The paper is organized as follows. In Section~\ref{2}, we introduce the model. The minisuperspace dynamics are presented in Subsection~\ref{2.1}, while the coupling to cosmological perturbations is discussed in Subsection~\ref{2.2}. The derivation of the master equation is given in Subsection~\ref{2.3}. In Subsection~\ref{2.4}, we discuss representative estimates of the relevant quantities and the corresponding decoherence timescale. Finally, Section~\ref{3} summarizes the results and discusses possible limitations and future directions.

\section{The model}
\label{2}

\subsection{Minisuperspace}
\label{2.1}

Consider a model with a single scalar inflaton
\cite{baumann2012tasilecturesinflation} and a clock field
\cite{Husain_2012}. The total action can be written as
\begin{equation}\label{action}
S[g_{\mu \nu}, \varphi, t, M] =  \int d^4 x \sqrt{-g}\, \left(\frac{M^2_{\text{Pl}}}{2} R -   \frac{1}{2} g^{\mu \nu} \partial_{\mu} \varphi \partial_{\nu} \varphi  - V(\varphi)-  \frac{M}{2} \left(g^{\mu \nu} \partial_{\mu} t \partial_{\nu} t +1 \right)\right).
\end{equation}
Here $g_{\mu \nu}$ denotes the metric\footnote{In this paper, the space-time indices $\mu, \nu, \lambda \dots$ run through the values $0, 1, 2, 3$ and the spatial indices $i, j, k\dots$ run through the values $ 1, 2, 3$.}, $g$ is its determinant, and $R$ is the scalar curvature. The variable $t$ denotes the clock field, $M$ is a Lagrange multiplier, $\varphi$ is the inflaton field, $V(\varphi)$ is its potential, and $M_{\text{Pl}}$ is the reduced Planck mass.

Using the ADM decomposition of the metric
$ds^2=- N^2 d \tau^2+ \gamma_{ik}(dx^i+ N^id \tau)(dx^k+ N^kd \tau)$ and following the procedure described in \cite{Husain_2012},
one can solve the equations of motion explicitly for $M$ and rewrite the dynamics in terms of the scalar and spatial diffeomorphism constraints $(\mathcal{H}_{\perp}, \mathcal{H}_{i})=0$. Imposing the gauge $t=\tau$ with $N=1$, one obtains
\begin{equation} \label{constraint}
\mathcal{H}_{\perp}=\pi_t + \underbrace{\mathcal{H}_{G} + \mathcal{H}_{M}}_{\mathcal{H}_{\text{phys}}}= 0,
\end{equation}
where $\mathcal{H}_{M}$ is the Hamiltonian density of the matter sector, corresponding in our case to the inflaton action, while $\mathcal{H}_{G}$ is the gravitational scalar constraint and $\pi_t$ is the canonical momentum conjugate to $t$. Thus, within this description, the clock field provides a time variable $t$ and $\mathcal{H}_{\text{phys}}$ plays the role of the corresponding physical Hamiltonian.

We are interested in the spatially flat cosmological solution, written as
\begin{equation} \label{metric}
ds^2=-d \tau^2+a^2(\tau) \delta_{ij}dx^{i} dx^j,\,\,\,\varphi=\varphi(\tau)
\end{equation}
where $a$ is the scale factor, $x^{i}$ are the spatial coordinates, and $\tau$ is the cosmological time. As is well known, in the slow-roll regime of the inflaton dynamics, the kinetic contribution is assumed to be subdominant,
$V \gg  \frac{1}{2}(\partial_{\tau} \varphi)^2$,
so that the contribution of the inflaton to the background gravitational dynamics can be approximated by an effective cosmological constant
\begin{equation}
\Lambda_{\varphi}=M^{-2}_{\text{Pl}} V (\varphi).
\end{equation}
This corresponds to the Hubble parameter
$H=\frac{\partial_{\tau} a}{a}=\sqrt{\Lambda_{\varphi} /3}$.
In the discussion of the unperturbed background in this section, we adopt this approximation within the framework of (\ref{metric}). The corresponding minisuperspace scalar constraint $\mathcal{H}_{\perp}=0$ then takes the form
\begin{equation} \label{scalar}
\mathcal{H}_{\perp} \approx \pi_{t}+ \frac{\pi^2_a}{24a}-2\Lambda_{\varphi} a^3,
\end{equation}
where $\pi_a$ is the canonical momentum conjugate to $a$.

It is important to note that the ordinary minisuperspace reduction leading to (\ref{scalar}) involves the spatial integral over the coordinate volume $\text{Vol}_x = \int d^3x$  and the factor $\frac{M^2_{\text{Pl}}}{2}$ from the action (\ref{action}) which were absorbed into the variable $a$, as
$\text{Vol}_x \frac{M^2_{\text{Pl}}}{2} a^3 \to a^3$.

As shown in \cite{Ali_2018}, performing the variable change
\begin{equation} \label{v}
x= \frac{4}{\sqrt{3}} a^{3/2},\,\,\, \Omega=\frac{3}{8}\Lambda_{\varphi},
\end{equation}
the constraint can be written in the simplified form
$\mathcal{H}_{\perp}= \pi_t+\frac{p^2}{2}-\Omega x^2$,
where $p$ is the canonical momentum conjugate to $x$. We emphasize that, within the present minisuperspace construction, $x$ is treated as a physical variable, with $[x]=M^{-1/2}_{\text{Pl}}$
\footnote{
The initial action is invariant under the transformation $x^i \to \lambda x^i, \quad a \to \lambda^{-1} a$, where $\lambda$ is a constant. However, this freedom is fixed after applying the minisuperspace reduction and obtaining (\ref{scalar}), where the spatial integrals and the constant $\frac{M^2_{\text{Pl}}}{2}$ are absorbed. This procedure leads to fixed physical variables $a$ and $x$. The relation $[x]=M^{-1/2}_{\text{Pl}}$ is consistent with $[t]=M^{-1}_{\text{Pl}}$.
}. Canonical quantization, with $p \to \hat{p}$ and $x \to \hat{x}$, then gives
\begin{equation}
i\frac{\partial}{\partial t} \Psi(x,t) =  \hat{\mathcal{H}}_{\text{phys}}\Psi(x,t) ,\,\,\, \hat{\mathcal{H}}_{\text{phys}}= \frac{\hat{p}^2}{2}-\Omega \hat{x}^2.
\end{equation}
Thus, within this description, the minisuperspace dynamics are equivalent to those of a harmonic oscillator with an inverted potential.

It is also useful to switch to the Heisenberg picture, in which a quantum operator $\hat{A}$ associated with a physical quantity evolves according to
\begin{equation}
\hat{A}(t) =  \exp\left(i t\left(\frac{\hat{p}^2}{2}-\Omega \hat{x}^2\right)\right) \hat{A}  \exp\left(-i t\left(\frac{\hat{p}^2}{2}-\Omega \hat{x}^2\right)\right).
\end{equation}
In particular, for $\hat{x}(t)$ one obtains
\begin{equation}\label{evolution}
\hat{x}(s)=\hat{x} \cosh (\sqrt{2 \Omega} s)+\frac{\hat{p}}{ \sqrt{2 \Omega}} \sinh (\sqrt{2 \Omega} s).
\end{equation}

We restrict the present model to the effective semiclassical regime in which the WKB approximation is expected to be valid, corresponding to sufficiently large values of $x$. The emergence of such a semiclassical regime is compatible with the characteristic energy scale of inflation being well below the Planck scale. More precisely, in terms of the WKB factor, this condition can be expressed as $\left|\frac{d}{dx} p^{-1} \right| \ll 1$. Its asymptotic estimate for $x \to \infty$ gives
\begin{equation}\label{WKB}
W=\left|\frac{d}{dx} p^{-1} \right|\approx \frac{1}{\sqrt{2}\Omega^{1/2} x^2} \ll 1 \quad \text{or}  \quad x \gg \Omega^{-1/4}.
\end{equation}
The corresponding classical slow-roll inflationary regime\footnote{It corresponds to the branch of solutions with the growing exponential ${x}(s)=x(0) \cosh (\sqrt{2 \Omega} s)+\frac{{p}(0)}{ \sqrt{2 \Omega}} \sinh (\sqrt{2 \Omega} s)$.} can then be written as
\begin{equation}\label{evol} 
x(t) = x(0) \exp\left(\sqrt{2 \Omega}t \right), \quad H= \frac{2\sqrt{2\Omega}}{3}.
\end{equation}

\subsection{Perturbations}
\label{2.2}

We consider the scalar sector of cosmological perturbations of the metric and the scalar field (see \cite{baumann2012tasilecturesinflation}) on the classical background given by (\ref{metric}), within the slow-roll approximation.  In the Newtonian gauge, the ansatz can be written as
\begin{equation} 
ds^2=-(1+2 \phi)d \tau^2+a^2(\tau) (1-2 \phi)\delta_{ij}dx^{i} dx^j,\,\,\, \tilde{\varphi}=\varphi(\tau)+\delta \varphi,
\end{equation}
where $\phi$ is the Bardeen potential, $\tilde{\varphi}$ denotes the full inflaton field, and $\delta \varphi$ is its perturbation.

We treat the dust as a test kinematical clock with negligible energy–momentum, so that the usual dynamics applies (see \cite{Giesel_2020}), and we identify the clock time with the longitudinal time $\tau$. 

We use the definition of the Mukhanov-Sasaki variable, written as $u=a \delta \varphi + a \frac{\partial_{\tau}\varphi}{H} \phi$. The corresponding quantum operator in the Heisenberg picture can be written as
\begin{equation}  \label{sol1}
\hat{u}(\mathbf{x}, \tilde{\tau})=\int \frac{d^3 k}{(2 \pi)^3}\left(v_k(\tilde{\tau}) \hat{a}_{\mathbf{k}} e^{i \mathbf{k x}}+v_k^*(\tilde{\tau}) \hat{a}_{\mathbf{k}}^{\dagger} e^{-i \mathbf{k x}}\right),
\end{equation}
where $\hat{a}_{\mathbf{k}}^{\dagger}$ and $\hat{a}_{\mathbf{k}}$ are the creation and annihilation operators, respectively, and
\begin{equation} \label{sol}
v_k=\frac{e^{-i k \tilde{\tau}}}{\sqrt{2 k}}\left(1-\frac{i}{k \tilde{\tau}}\right),
\end{equation}
where $\tilde{\tau}$ is the conformal time, $\tilde{\tau} = -\frac{1}{aH}$, and $k$ denotes the magnitude of the wave vector $\mathbf{k}$. We use notation 
$[\hat{a}_{\mathbf{k}}, \hat{a}^{\dagger}_{\mathbf{p}}]=(2\pi)^3 \delta^{(3)} (\mathbf{k}-\mathbf{p})$.

Following the phenomenological scenario considered in \cite{li2019preinflationaryperturbationsthermalbackground}, we assume that the perturbations are described by a thermal state with density matrix $\hat{\rho}_{\mathcal{T}}$, defined through
\begin{equation} \label{def}
\langle \hat{a}_{\mathbf{k}}^{\dagger} \hat{a}_{\mathbf{p}}\rangle_{\mathcal{T}} = \operatorname{Tr} [ \hat{\rho}_{\mathcal{T}} \hat{a}_{\mathbf{k}}^{\dagger} \hat{a}_{\mathbf{p}} ] = (2\pi)^3 \delta^{(3)}(\mathbf{k}-\mathbf{p}) \frac{1}{e^{\frac{k}{\mathcal{T}}}-1},
\end{equation}
with $\mathcal{T}$ denoting an effective (comoving) temperature. One can also define the corresponding physical temperature as $\mathcal{T}_{\text{phys}}= \frac{\mathcal{T}}{a}$.

We stress that the thermal character of the perturbations is an assumption of the present toy model rather than a result derived from the underlying cosmological dynamics. In particular, we do not attempt to derive the initial state of the perturbations from a specific mechanism of pre-inflationary dynamics, or quantum gravity. The purpose of this assumption is to provide a concrete and tractable example in which the perturbations can be treated as an environment within the open quantum systems framework. Alternative choices of the initial state could lead to quantitatively or even qualitatively different results and therefore constitute a natural subject for further investigation.

We consider the low-temperature regime $\mathcal{T} \ll aH$. Then, because of the suppression of the thermal occupation factors by terms of the form $\sim e^{-\frac{k}{\mathcal{T}}}$, modes with $k \gg \mathcal{T}$ are exponentially suppressed in the thermal contribution. Consequently, within the assumptions of the present model, the dominant contribution comes from the modes which are in the super-horizon regime. For (\ref{sol}), this corresponds to the limit
\begin{equation}
v_k \to - \frac{i}{\sqrt{2}} \frac{1}{k^{3/2}\tilde{\tau}}.
\end{equation}

In what follows, we consider only super-horizon modes. This restriction should be understood as conditional on the assumed hierarchy of scales and on the thermal state ansatz introduced above.

\subsection{Master equation}
\label{2.3}
We now introduce an interaction between the cosmological sector and the perturbations. We consider the backreaction approach as an effective description that is compatible with the assumption of a weak influence of the system on its environment. Within this approach, the equations governing the cosmological perturbations remain unchanged to the order considered, while the cosmological sector acquires additional effective contributions \cite{Mukhanov_1997, li2008cosmological}.

More precisely, the Einstein equations in the form $P_{\mu \nu}=G_{\mu \nu}-M^2_{\text{Pl}}T_{\mu \nu}=0$ are expanded in terms of the variables $q^{\alpha}=(g_{\mu \nu }, \varphi)$, and the effective spatial averaging $<>$ is applied (so that the cosmological solution is regarded as an emergent description of the background\footnote{Averages of quantities at the expansion point (which is treated as the background) are effectively taken as the corresponding quantities at the cosmological solution.}) as
\begin{equation} 
P_{\mu \nu}\left(q^\alpha\right)+ <\underbrace{\left.\frac{\delta P_{\mu \nu}\left(q^{\prime \alpha}\right)}{\delta q^{\prime \alpha}}\right|_{q^{\prime \alpha}=q^\alpha} \delta q^\alpha}_{=0}>+ \underbrace{\left.\frac{\delta^2 P_{\mu \nu}\left(q^{\prime \alpha}\right)}{\delta q^{\prime \alpha} \delta q^{\prime \beta}}\right|_{q^{\prime \alpha}=q^\alpha}< \delta q^\alpha \delta q^\beta>}_{ -M^{2}_{\text{Pl}}T_{ \mu \nu\,\text{pert}}}=0.
\end{equation}
Here, the second term vanishes on the equations of motion for the perturbations, while the last term represents the contribution of the effective stress-energy tensor $T_{\mu \nu\,\text{pert} }$.

It is known \cite{Abramo_1997, brandenberger2002reactioncosmologicalperturbationscosmological} that the backreaction of the scalar super-horizon modes results in the effective cosmological constant 
\begin{equation} 
\Lambda_{\text{pert}}=M^{-2}_{\text{Pl}}\left(2 \frac{V^{\prime \prime}(\varphi) V^2(\varphi)}{V^{\prime 2}(\varphi)}-4 V(\varphi)\right)<\phi^2>,
\end{equation}
which can be applied to our case\footnote{As was stated, we restrict the toy model to pure scalar perturbations of the metric ansatz and the inflaton. In particular, we also do not take into account possible clock perturbations associated with the initial synchronization (see \cite{Geshnizjani_2005}).
}. Note that the physical interpretation of this effect depends on the choice of the physical clock, see \cite{unruh1998cosmologicallongwavelengthperturbations, Geshnizjani_2002}.

We introduce the corresponding quantum operator $\hat{\Lambda}_{\text{pert}}$. It can be written, using the expansion of $u$ in terms of $v_k$ and its quantization 
(\ref{sol1}), as 
\begin{equation}  \label{pert}
\hat{\Lambda}_{\text{pert}} \approx  \frac{16}{27 } M^{-2}_{\text{Pl}} (\eta-4\epsilon ) \Omega^2  \frac{1}{ \text{Vol}_{x}}\mathop{\int} \limits_{\!{k>k_{\text{IR}}}} \frac{d^3\mathbf{k} }{(2 \pi)^{3} } \frac{1}{k^{3}}   (\hat{a}^{\dagger}_{\mathbf{k}}\hat{a}_{\mathbf{k}}+\hat{a}_{\mathbf{k}}\hat{a}^{\dagger}_{\mathbf{k}} -\hat{a}^{\dagger}_{\mathbf{k}}\hat{a}^{\dagger}_{-{\mathbf{k}}}-\hat{a}_{\mathbf{k}}\hat{a}_{-{\mathbf{k}}} )
\end{equation}
with the slow-roll parameters
$\epsilon = \frac{M^2_{\text{Pl}}}{2} \left(\frac{V'(\varphi)}{V(\varphi)}\right)^2$ and $
\eta = {M^2_{\text{Pl}}} \frac{V''(\varphi)}{V(\varphi)}$, where $\epsilon \ll 1$ is assumed in obtaining the approximate expression. The quantity $k_{\text{IR}}$ denotes infrared cutoff, and $\text{Vol}_{x}$ is the coordinate spatial volume introduced by the spatial averaging. The physical interpretation and choice of these regularizations will be discussed in the context of the estimates in the next subsection.

Within the presented framework, we work in the interaction picture. The evolution of the system associated with (\ref{evolution}) is taken into account, while the time dependence of the perturbations is already encoded in the mode functions $v_k$. We therefore write the interaction Hamiltonian as the quantization of the effective cosmological-constant contribution:
\begin{equation} 
\hat{\mathcal{H}}_{\text{int}} (t)=  - \frac{3}{8} \hat{\Lambda}_{\text{pert}} \hat{x}^2(t).
\end{equation}

We follow the standard second-order perturbative derivation of a time-local master equation (see \cite{Manzano_2020}), while deliberately not imposing the Markovian limit itself. Namely, we write the interaction Hamiltonian in the form
$\hat{\mathcal{H}}_{\text{int}} (t)= \sum_{\alpha} \mathcal{S}_{\alpha}(t) \otimes \mathcal{B}_{\alpha}(t) $,
where, in the present case, the bath operator is $\mathcal{B}_{1}= \frac{3}{8} \hat{\Lambda}_{\text{pert}} $ and the system operator is $\mathcal{S}_{1}=\hat{x}^2(t)$. We assume the Born approximation for the total density matrix $\hat{\rho}_{\text{tot}}$ in the form
$\hat{\rho}_{\text{tot}}(t)= \hat{\rho}_{\mathcal{T}} \otimes \hat{\rho}(t)$
up to the order retained in the perturbative treatment, where $\hat{\rho}$ is the reduced density matrix of the system and $\hat{\rho}_{\mathcal{T}}$ is the thermal state assumed for the environment. This factorized form should be regarded as a working approximation of the present toy model. In particular, we do not establish its validity from an underlying cosmological dynamics. After tracing out the bath degrees of freedom, we obtain the following expression. Let us briefly outline the relevant steps.

We begin with the dynamics in the interaction representation, indicated by the superscript $\text{I}$ in the density matrices,
\begin{equation}
\partial_t \hat{\rho}^{\text{I}}_{\text{tot}}(t)= -i[\hat{\mathcal{H}}_{\text{int}} (t), \hat{\rho}^{\text{I}}_{\text{tot}}(t)]
\end{equation}
and substitute back the solution $\hat{\rho}^{\text{I}}_{\text{tot}}(t)=\hat{\rho}^{\text{I}}_{\text{tot}}(0) -i\int^t_0 ds[\hat{\mathcal{H}}_{\text{int}} (s), \hat{\rho}^{\text{I}}_{\text{tot}}(s)]$:
\begin{equation}
\partial_t \hat{\rho}^{\text{I}}_{\text{tot}}(t)= -i[\hat{\mathcal{H}}_{\text{int}} (t), \hat{\rho}^{\text{I}}_{\text{tot}}(0)]-\int^t
_0 ds[\hat{\mathcal{H}}_{\text{int}} (t),[\hat{\mathcal{H}}_{\text{int}} (s), \hat{\rho}^{\text{I}}_{\text{tot}}(s)]].
\end{equation}
Next, inserting the Born approximation as $\hat{\rho}^{\text{I}}_{\text{tot}}(t)= \hat{\rho}_{\mathcal{T}}^{\text{I}} \otimes \hat{\rho}^{\text{I}}(t)$ and applying the trace of the bath degrees of freedom with $\operatorname{Tr}_{\mathcal{B}}$, one obtains
\begin{equation}
\partial_t \hat{\rho}^{\text{I}}(t)= -i\operatorname{Tr}_{\mathcal{B}}[\hat{\mathcal{H}}_{\text{int}} (t), \hat{\rho}^{\text{I}}_{\text{tot}}(0)]-\int^t
_0 ds\operatorname{Tr}_{\mathcal{B}}[\hat{\mathcal{H}}_{\text{int}} (t),[\hat{\mathcal{H}}_{\text{int}} (s), \hat{\rho}^{\text{I}}_{\text{tot}}(s)]].
\end{equation}
or
\begin{equation}
\partial_t \hat{\rho}^{\text{I}}(t)=-i \sum_\alpha\left\langle\hat{\mathcal{B}}_\alpha(t)\right\rangle_{\mathcal{T}}  {\left[\hat{\mathcal{S}}_\alpha(t), \hat{\rho}^{\text{I}}(0)\right]-\sum_{\alpha, \beta} \int_0^t \frac{d s}{2}\left(\left\langle\left\{\hat{\mathcal{B}}_\alpha(t), \hat{\mathcal{B}}_\beta(s)\right\}\right\rangle_{\mathcal{T}} \times\right.} \nonumber
\end{equation}
\begin{equation}
\times\left[\hat{\mathcal{S}}_\alpha(t),\left[\hat{\mathcal{S}}_\beta(s),\hat{\rho}^{\text{I}}(t)\right]\right]\left.+\left\langle\left[\hat{\mathcal{B}}_\alpha(t), \hat{\mathcal{B}}_\beta(s)\right]\right\rangle_{\mathcal{T}}\left[\hat{\mathcal{S}}_\alpha(t),\left\{\hat{\mathcal{S}}_\beta(s), \hat{\rho}^{\text{I}}(t)\right\}\right]\right),
\end{equation}
up to the next order. Finally, inserting the relation $\hat{\rho}^{\text{I}}_{\text{tot}}(0)=\hat{\rho}^{\text{I}}_{\text{tot}}(t) +i\int^t_0 ds[\hat{\mathcal{H}}_{\text{int}} (s), \hat{\rho}^{\text{I}}_{\text{tot}}(s)]$, performing the reparameterization $s \to t-s$ in the integrals, one obtains the following form of the master equation in the Schrödinger picture:
\begin{equation}
\partial_t  \hat{\rho} (t) = \mathcal{L} \hat{\rho}= -i[\hat{\mathcal{H}_S}, \hat{{\rho}}(t) ]  -i   \sum_{\alpha} \langle \hat{\mathcal{B}}_{\alpha}(t) \rangle_{\mathcal{T}} \left[\hat{\mathcal{S}}_{\alpha}(t), \hat{\rho}(t)   \right]  
 +  \sum_{\alpha, \beta} \int^{t}_0 ds     \langle \hat{\mathcal{B}}_{\alpha}(t) \rangle_{\mathcal{T}}   \langle \hat{\mathcal{B}}_{\beta}(t-s) \rangle_{\mathcal{T}} \times
\nonumber
\end{equation}
\begin{equation}
 \times\left[\hat{\mathcal{S}}_{\alpha}(t), \left[\hat{\mathcal{S}}_{\beta}(t-s), \hat{\rho}(t)   \right]   \right]  
-\sum_{\alpha,\beta}\int_0^{t}  \frac{ds}{2}\left(\langle{ \{\hat{\mathcal{B}}_{\alpha}(t), \hat{\mathcal{B}}_{\beta}(t-s)\}} \rangle_{\mathcal{T}}[\hat{\mathcal{S}}_{\alpha}(t),[\hat{\mathcal{S}}_{\beta}(t-s),\hat{\rho}(t)]] \right.   \nonumber
\end{equation}
\begin{equation}\label{me}
+  \left. \langle{ [\hat{\mathcal{B}}_{\alpha}(t), \hat{\mathcal{B}}_{\beta}(t-s)]} \rangle_{\mathcal{T}}[\hat{\mathcal{S}}_{\alpha}(t),\{\hat{\mathcal{S}}_{\beta}(t-s),\hat{\rho}(t)\}]\right),
\end{equation}
where $\mathcal{L}$ is the Liouvillian superoperator.

For our case, this form of the master equation can be written as
\begin{equation} 
\partial_{t} \hat{\rho}= -i[\frac{\hat{p}^2}{2}-\Omega \hat{x}^2,\hat{\rho}]  +i \frac{3}{8}\langle \hat{\Lambda}_{\text{pert}} \rangle_{\mathcal{T}} [\hat{x}^2, \hat{\rho}] 
+\frac{9}{64}(\langle \hat{\Lambda}_{\text{pert}} \rangle_{\mathcal{T}} \langle \hat{\Lambda}_{\text{pert}} \rangle_{\mathcal{T}} - \langle \hat{\Lambda}_{\text{pert}} \hat{\Lambda}_{\text{pert}}\rangle_{\mathcal{T}})  \times \nonumber
\end{equation} 
\begin{equation}\label{me1}\times(A_1(t)[\hat{x}^2,[\hat{x}^2,\hat{\rho}]]+A_2(t)[\hat{x}^2,[\hat{p}^2,\hat{\rho}]]+A_3(t)[\hat{x}^2,[\hat{x}\hat{p},\hat{\rho}]]).
\end{equation}
Note that $t$ corresponds to the duration of the considered regime. The $t$-dependent functions are:
\begin{equation}
A_1(t) = \frac{2 \sqrt{2 \Omega} t  + \sinh(2 \sqrt{2 \Omega} t )}{4 \sqrt{2 \Omega}},\quad A_2(t)= \frac{-2 \sqrt{2 \Omega} t  + \sinh(2 \sqrt{2 \Omega} t )}{8 \Omega \sqrt{2 \Omega}}, \quad A_3(t) = \frac{\sinh^2(\sqrt{2 \Omega} t )} {2 \Omega}.
\end{equation}
It is important to note that the Markovian approximation cannot be applied in the usual form, which would require the limit $t\to \infty$ together with sufficiently localized bath correlation functions in time. In the present model, the coefficients $A_1$, $A_2$, and $A_3$ exhibit exponential behavior in this limit, preventing such a simplification from being straightforwardly justified.

The first term in (\ref{me1}) corresponds to the usual unitary evolution in the Schrödinger picture, while the second term can be interpreted as a generalized Lamb shift and can be absorbed into $\Omega$ by the renormalization $\Omega \to \Omega + \frac{3}{8}\langle \hat{\Lambda}_{\text{pert}} \rangle_{\mathcal{T}}$, up to the order. The structure of the next term, $\sim [\hat{x}^2,[\hat{x}^2,\hat{\rho}]]$, is similar to the decoherence term appearing in the Caldeira-Leggett model at high temperatures \cite{CALDEIRA1983587, zurek2003decoherencetransitionquantumclassical}, with the first power of the coordinate replaced by the second power. This term is non-unitary and can therefore contribute to the suppression of quantum coherence.  The remaining terms, which have a less standard structure, are not leading compared with the first coupling term in the very fast regime considered below. Namely, we consider the regime corresponding to the small duration as a relative e-folding number limit $N_{exp}= \frac{2}{3} \sqrt{2\Omega} t \to 0$. In addition,  we remove the vacuum contribution in $ \langle\hat{\Lambda}_{\text{pert}}\rangle_{\mathcal{T}}$ itself through normal ordering $:\hat{\Lambda}_{\text{pert}}:$. Thus, we write, to leading order in the Taylor expansion:
\begin{equation}  \label{mef}
\partial_{t} \hat{\rho} \approx -i[\frac{\hat{p}^2}{2}-\Omega \hat{x}^2,\hat{\rho}]  
-\underbrace{\frac{9}{64}(  \langle: \hat{\Lambda}_{\text{pert}}: : \hat{\Lambda}_{\text{pert}}:\rangle_{\mathcal{T}}-\langle: \hat{\Lambda}_{\text{pert}}: \rangle_{\mathcal{T}} \langle: \hat{\Lambda}_{\text{pert}}: \rangle_{\mathcal{T}})}_{\lambda} t [\hat{x}^2,[\hat{x}^2,\hat{\rho}]].
\end{equation} 

To determine the coupling constant $\lambda$ in (\ref{mef}), one needs to calculate the thermal correlation functions of the various products of creation and annihilation operators arising from (\ref{pert}). These correlation functions can be evaluated using (\ref{def}) together with the Kubo-Martin-Schwinger condition and commutation relations. One obtains the integral,  using spherical coordinates in momentum space:
\begin{equation} 
\lambda \approx \underbrace{\frac{4}{81 \pi^2} M^{-4}_{\text{Pl}} (4 \epsilon - \eta)^2 \Omega^4
 \frac{1}{\text{Vol}_x } \left[ \mathop{\int} \limits^{\infty }_{k_{\text{IR}}}  \frac{dk }{k^4 }  (4n^2_k+4n_k) \right.}_{\lambda_{\mathcal{T}}} + \underbrace{ \left. \mathop{\int} \limits^{\infty }_{k_{\text{IR}}}  \frac{dk }{k^4 } \right]}_{\text{vacuum term}},
\end{equation} 
where $n_k= \frac{1}{e^{\frac{k}{\mathcal{T}}}-1}$. Being interested in the thermal effect, we do not analyze the vacuum contribution here\footnote{The vacuum contribution requires a separate treatment, where the key aspect is its regularization and renormalization. In general, the argument of the suppression of the sub-horizon modes is not applicable here. Sub-horizon modes must be taken into account, which will lead to additional contributions from the back-reaction and the ultraviolet problem.}. We consider only the thermal contribution in this toy model.

Finally, in the leading order, using spherical coordinates, our coupling constant is given by

\begin{equation} 
\lambda_{\mathcal{T}} \approx \frac{16}{81 \pi^2} M^{-4}_{\text{Pl}} (4 \epsilon - \eta)^2 \Omega^4
 \frac{1}{\text{Vol}_x }  \int\limits^{\infty }_{k_{\text{IR}}} \frac{d k  }{k^4} e^{-\frac{k}{\mathcal{T}}}.
\end{equation} 

\subsection{Numerical estimates}
\label{2.4}

Having formulated the model, we now turn to estimating the parameters that characterize the potential effect. First, note that within the backreaction formalism, cosmological parameters are emergent quantities obtained by averaging over a certain spatial region, and this dependence is explicitly reflected in the corresponding quantities, see \cite{li2008cosmological}. By analogy with \cite{brandenberger2002reactioncosmologicalperturbationscosmological}, we take the infrared cutoff as the longest wavelength corresponding to the beginning of inflation, since the longer wavelengths are not related to the inflationary cosmology. We set the corresponding volume regularization as the initial Hubble sphere.

Namely, we set $k_{\text{IR}}= H a_i$, where $a_i$ is the initial scale factor interpreted as corresponding to the stage at which the thermal generalization of the Bunch-Davies vacuum was prepared. For the volume regularization we set $\text{Vol}_x=\frac{4 \pi}{3}  (a_{\text{i}} H)^{-3}$. We also emphasize that we do not consider the ultraviolet regularization in the integral, due to the suppression of high-frequency modes by the temperature distribution. This gives, to leading order:
\begin{equation} 
\lambda_{\mathcal{T}}  \approx \frac{4}{27  \pi^{3} }  M^{-4}_{\text{Pl}} (4 \epsilon - \eta)^2 \Omega^4 
  (Ha_{\text{i}})^3 \int\limits^{\infty}_{Ha_{\text{i}}} \frac{dk  }{k^{4} }    e^{-\frac{k}{\mathcal{T}}}.
\end{equation}
Introducing the dimensionless parameter $\tilde{\mathcal{T}} =\mathcal{T} (Ha_{\text{i}})^{-1}$, the equation can be written as 
\begin{equation} 
\lambda_{\mathcal{T}}  \approx \frac{4}{27 \pi^3} M_{\text{Pl}}^{-4}(4 \epsilon-\eta)^2 \Omega^4   \tilde{\mathcal{T}}^{-3} \Gamma\left(-3, \frac{1}{\tilde{\mathcal{T}}}\right),
\end{equation}
where $\Gamma(s,t) $ denotes the incomplete Gamma function.

We are interested in the decoherence process associated with the second term in the master equation
\begin{equation} 
\label{meft}
\partial_{t} \hat{\rho} \approx -i[\frac{\hat{p}^2}{2}-\Omega \hat{x}^2,\hat{\rho}]  
-\lambda_{\mathcal{T}} t [\hat{x}^2,[\hat{x}^2,\hat{\rho}]].
\end{equation}
In the coordinate representation, this term leads to an exponential suppression of the off-diagonal elements of the density matrix of the form $\sim e^{- \lambda_{\mathcal{T}} \frac{t^2}{2}(x'^2-x^2)^2 }$. The corresponding characteristic decoherence timescale for quantum correlations at the scale of the universe $x$, for which we set $(x'^2-x^2)^2 \sim x^4$, can be defined as
\begin{equation}
t_{\text{dec}} = \sqrt{2}\lambda_{\mathcal{T}} ^{-1/2} x^{-2}.
\end{equation}

Using $W$ from (\ref{WKB}), the definition $\Omega=\frac{9}{8} H^2$ and applying $\tilde{T} \ll 1$, one can estimate, in the leading order:
\begin{equation}
t_{\text{dec}} \approx 24 M^{2}_{\text{Pl}} W \frac{\exp{\left(\frac{1}{2}\tilde{T}^{-1}\right)}}{\sqrt{\tilde{T}}} |4\epsilon-\eta|^{-1} H^{-3}.
\end{equation}
This relation provides an estimate of the timescale associated with the suppression of the quantum cosmological correlations and, consequently, can be used as an indicative measure of the induced quantum-to-classical transition. In particular, inequality $t_{\text{dec}} \ll H^{-1}$ sets the restrictions of the corresponding parameters.

As an illustrative estimate of the relevant scales, let us consider  $\tilde{\mathcal{T}}=0.1$ (taking into account the limitations of the temperature from
 \cite{li2019preinflationaryperturbationsthermalbackground}). In accordance with characteristic values in inflationary models, we set $|4\epsilon-\eta| \approx 10^{-2}$ 
and take the characteristic inflationary scale, namely $V(\varphi)\approx 10^{-10}M^{4}_{\text{Pl}}$, so $H \approx 10^{-5} M_{\text{Pl}} $ (see, for example, \cite{Liddle_2003, refId0, baumann2012tasilecturesinflation}). Then, one will get
\begin{equation}\label{time}
t_{\text{dec}} \ll H^{-1}\,\,\,\text{for}\,\,\, W \ll 10^{-16}.
\end{equation}

To illustrate the possibility of such a scale of $W$, recall that our minisuperspace variable $x$ consists of $ \sim  M_{\text{Pl}} \sqrt{\text{Vol}_x a^3}$ ($a$ here is the initial scale factor with gauge freedom). Substituting our regularized volume $\text{Vol}_x = \frac{4\pi}{3} (Ha_i)^{-3}$ into the definition of $x$, one can get for the WKB parameter:
\begin{equation}
W \approx \frac{3}{16 \pi}  \frac{H^2}{M^2_{\text{Pl}}} \left(\frac{a_i}{a}\right)^3.
\end{equation}
Introducing the corresponding e-folding number $N_0=\ln\left( \frac{a}{a_i} \right)$, we get that $W \approx10^{-16}$ (for which $t_{\text{dec}} H=1$) corresponds to $N_0 \approx 3.8$.
Thus, the question concerns the applicability of the model at scales on the order of several e-foldings. In this note, we treat the model as phenomenological and do not delve into its implementation. However, we offer the technical observation that the infrared cutoff $k_{\text{IR}} = a_i H$ strongly suppresses modes at low $N_0 < 1$. The environment is expected to be formed at the scales of our estimate.

In addition, let us estimate the thermal backreaction itself given by
\begin{equation}
\langle \hat{\Lambda}_{\text{pert}} \rangle_{\mathcal{T}} = \frac{3 } {4 \pi^2} M^{-2}_{\text{Pl}} (\eta - 4\epsilon) H^4 \int\limits^{\infty}_{Ha_{\text{i}}} \frac{dk  }{k }   e^{-\frac{k}{\tilde{\mathcal{T}} a_i H}} =\frac{3 } {4 \pi^2} M^{-2}_{\text{Pl}} (\eta - 4\epsilon) H^4 \Gamma\left(0, \frac{1}{\tilde{\mathcal{T}}}\right) .
\end{equation}
For the same parameter range we get 
\begin{equation}
|\langle \hat{\Lambda}_{\text{pert}} \rangle_{\mathcal{T}}| \approx 10^{-29} M^{2}_{\text{Pl}} \ll    M^{-2}_{\text{Pl}} V(\varphi).
\end{equation}
While, for the coupling itself:
\begin{equation}
|\lambda_{\mathcal{T}} ^{1/2}| \approx 10^{-26} M^{2}_{\text{Pl}} \ll    M^{-2}_{\text{Pl}} V(\varphi).
\end{equation}
This highlights that the rapid decoherence observed here is a consequence not of strong interaction, but of the semiclassical nature of the cosmological system.

This estimate suggests that, for the particular parameter choices considered above, the decoherence timescale may be substantially shorter than the characteristic expansion timescale. If this hierarchy holds, the regime described by our model occurs very rapidly. It would provide support for the applicability of the assumptions made for the derivation of (\ref{meft}). In particular, the corresponding limit $N_{exp}= \frac{2}{3}\ln\left(\frac{x(t)}{x(0)}\right) \to 0$ is then consistent with the semiclassical short-time treatment. Moreover, the system Hamiltonian remains a minor contribution over such short times for the considered cosmological scales, since $\hat{x}(t_{\text{dec}}) \approx \hat{x}(0)$. Thus, within these approximations, one can write the cosmological density matrix solution $\rho(x',x,t)$ in the form
\begin{equation} \label{deco}
\rho(x',x,t) \approx \exp \left( - \lambda_{\mathcal{T}}  \frac{t^2}{2}(x'^2-x^2)^2 \right) \rho(x',x,0).
\end{equation}

Since the super-horizon modes are frozen, the bath correlation function is
time-independent, and (\ref{deco}) coincides with an average of the phase factor
$\exp[-i\tfrac{3}{8}\,\delta\Lambda_{\text{pert}}\, t\,(x'^2-x^2)]$ over Gaussian
fluctuations $\delta\Lambda_{\text{pert}}$ with variance
$\langle\hat{\Lambda}_{\text{ pert}}^2\rangle_{\mathcal{T}}-\langle\hat{\Lambda}_{\text{pert}}\rangle_{\mathcal{T}}^2$,
which explains the $t^2$ dependence of the exponent.

In the strong-suppression limit of (\ref{deco}), superpositions between states with different eigenvalues of $\hat{x}^2$ are strongly suppressed, which is suggestive of an effective classical description in terms of approximate eigenstates of $\hat{x}^2$.

More generally, according to the exponential factor in (\ref{deco}), the eigenstates of $\hat{x}^2$ can be regarded as candidate pointer states of the cosmological system, as they are the states most robust against the decoherence induced by the interaction considered here. Since the variable $x$ is restricted to positive values in the present model (see (\ref{v})), their difference from the eigenstates of $\hat{x}$ is not relevant in this consideration.

This selection corresponds to the quantum-measurement limit, in which the system Hamiltonian is negligible. When it competes with the interaction, the form of the pointer states may differ. In particular, they are coherent states for the harmonic oscillator with a non-unitary interaction of the form $\sim [\hat{x},[\hat{x},\hat{\rho}]]$
 \cite{PhysRevLett.70.1187}.

\section{Discussion}
\label{3}

We have considered a toy model of decoherence in quantum cosmology based on an effective interaction between the quantum degrees of freedom of a homogeneous inflationary universe and scalar metric and inflaton perturbations at finite temperature. The latter are treated as an environment in an open quantum systems framework, with their influence on the homogeneous sector described through the backreaction formalism. Under the assumptions adopted in the model, the reduced dynamics induced by the thermal contribution can be described, at sufficiently short times, by a master equation containing the non-unitary contribution $-\lambda_{\mathcal{T}}  t [\hat{x}^2,[\hat{x}^2,\hat{\rho}]]$,
where $x\sim a^{3/2}$ is the minisuperspace variable introduced above and $t$ is the cosmological time in terms of the clock field. 

The corresponding decoherence timescale was estimated to be parametrically shorter than the Hubble time for the representative set of parameters considered in Section \ref{2.4}. This estimate should, however, be regarded as illustrative, since it depends on the assumptions and the parameters, the main ones of which are the minisuperspace scale of the considered regime and the temperature. In future work, it is reasonable to investigate the stability of this conclusion.

Within the short-time approximation used in the analysis, on cosmological scales, where the contribution of the system Hamiltonian is small,
the reduced density matrix takes the approximate form
$\rho(x',x,t) \approx e^{-\lambda_{\mathcal{T}}  t^2 (x'^2-x^2)^2/2}\rho(x',x,0).$
This expression indicates a suppression of coherences between states with different values of $\hat{x}^2$. Since $x\sim a^{3/2}$, this can be effectively interpreted as a suppression of coherences between different values of the scale factor. In this sense, the dynamics considered here are consistent with an effective description in terms of approximately classical branches characterized by different values of the scale factor. This result is qualitatively related to earlier approaches to decoherence in quantum cosmology, such as \cite{Barvinsky_1998}, in which these classical branches emerge during the onset of inflation.

This observation suggests a possible connection with the discussion of semiclassical states, in particular in loop quantum cosmology. In LQC models coherent states are often employed in the description of semiclassical dynamics \cite{Taveras_2008} and criteria for semiclassical states are also under discussion \cite{Corichi_2011}. However, in the present model, the interaction structure singles out $\hat{x}^2$ as the operator whose eigenstates are least affected by the corresponding decoherence term. Such states may therefore be regarded as candidate pointer states within the restricted framework considered here. This observation, being a toy counterexample, could motivate further investigation of whether decoherence effects of this type in more comprehensive models may provide an additional criterion for the selection of semiclassical states in quantum cosmology.

However, it should also be noted that the distinction between decoherence due
to entanglement and ensemble dephasing is not
apparent at the level of the reduced density matrix. See \cite{Schlosshauer_2019} for the details. According to the results in Section \ref{2.4}, classical dephasing occurs in this model if the value of the effective thermal
backreaction is treated as a classical unknown. This question is therefore
directly related to the nature of the thermal state, the origin of which lies
beyond the scope of this note.

Several limitations of the framework should be kept in mind. The thermal state of the scalar perturbations is introduced phenomenologically and is not derived from a specific pre-inflationary mechanism. Likewise, the Born approximation, the backreaction description, the regularization, and the short-time expansion are working assumptions whose range of validity has not been established in detail here. The analysis also considers only thermal contribution from inflaton and metric sector and neglects possible effects associated with other modes. In general, these extensions could modify both the magnitude and the structure of the resulting decoherence effect.

Subject to these qualifications, the model provides an illustrative example of how cosmological perturbations in a non-vacuum state can be incorporated into an open-system description of quantum cosmological degrees of freedom. In particular, this may be significant when considering primordial cosmological perturbations in quantum cosmology at finite temperature. The results suggest that, under certain assumptions, the associated thermal backreaction may lead to a suppression of coherences in the scale-factor sector on timescales shorter than the characteristic expansion time, together with the emergence of approximate pointer-state branches.

\section*{Acknowledgments}
This note is a continuation of the ideas presented in the author's Master's thesis \cite{Kulyabin:2025MSc}, carried out at the Institute for Quantum Gravity, FAU Erlangen-Nürnberg. The author thanks Kristina Giesel for suggesting the direction of this research and for discussions. The author is also grateful to Renata Ferrero for discussions and comments.

\sloppy
\printbibliography

@article{Halliwell:1989vw,
    author = "Halliwell, Jonathan J.",
    title = "{Decoherence in Quantum Cosmology}",
    reportNumber = "NSF-ITP-89-07",
    doi = "10.1103/PhysRevD.39.2912",
    journal = "Phys. Rev. D",
    volume = "39",
    pages = "2912",
    year = "1989"
}

@article{Liddle_2003,
   title={How long before the end of inflation were observable perturbations produced?},
   volume={68},
   ISSN={1089-4918},
   url={http://dx.doi.org/10.1103/PhysRevD.68.103503},
   DOI={10.1103/physrevd.68.103503},
   number={10},
   journal={Physical Review D},
   publisher={American Physical Society (APS)},
   author={Liddle, Andrew R. and Leach, Samuel M.},
   year={2003},
   month=Nov }

@article{Barvinsky_1998,
   title={Decoherence in quantum cosmology at the onset of inflation},
   volume={551},
   ISSN={0550-3213},
   url={http://dx.doi.org/10.1016/S0550-3213(99)00208-4},
   DOI={10.1016/s0550-3213(99)00208-4},
   number={1-2},
   journal={Nuclear Physics B},
   publisher={Elsevier BV},
   author={Barvinsky, A.O and Kamenshchik, A.Yu and Kiefer, C and Mishakov, I.V},
   year={1999},
   month=June, pages={374–396}}

@article{PhysRevD.31.1777,
  title = {Origin of structure in the Universe},
  author = {Halliwell, J. J. and Hawking, S. W.},
  journal = {Phys. Rev. D},
  volume = {31},
  issue = {8},
  pages = {1777--1791},
  numpages = {0},
  year = {1985},
  month = {Apr},
  publisher = {American Physical Society},
  doi = {10.1103/PhysRevD.31.1777},
  url = {https://link.aps.org/doi/10.1103/PhysRevD.31.1777}
}

@misc{novikov2021quantumdecoherencehomogeneousmodes,
      title={Quantum decoherence of the homogeneous modes during inflation}, 
      author={Oleg O. Novikov and Arina M. Shtennikova},
      year={2021},
      eprint={2112.06894},
      archivePrefix={arXiv},
      primaryClass={gr-qc},
      url={https://arxiv.org/abs/2112.06894}, 
}

@inbook{Barvinsky_2022,
   title={Preferred Basis, Decoherence and the Quantum State of the Universe},
   ISBN={9783030887810},
   ISSN={2365-6425},
   url={http://dx.doi.org/10.1007/978-3-030-88781-0_6},
   DOI={10.1007/978-3-030-88781-0_6},
   booktitle={From Quantum to Classical},
   publisher={Springer International Publishing},
   author={Barvinsky, Andrei O. and Kamenshchik, Alexander Yu.},
   year={2022},
   pages={97–120} }

@article{Anastopoulos_2013,
   title={A master equation for gravitational decoherence: probing the textures of spacetime},
   volume={30},
   ISSN={1361-6382},
   url={http://dx.doi.org/10.1088/0264-9381/30/16/165007},
   DOI={10.1088/0264-9381/30/16/165007},
   number={16},
   journal={Classical and Quantum Gravity},
   publisher={IOP Publishing},
   author={Anastopoulos, C and Hu, B L},
   year={2013},
   month=July, pages={165007} }

@book{10.1093/acprof:oso/9780199213900.001.0001,
    author = {Breuer, Heinz-Peter and Petruccione, Francesco},
    title = {The Theory of Open Quantum Systems},
    publisher = {Oxford University Press},
    year = {2007},
    month = {01},
    isbn = {9780199213900},
    doi = {10.1093/acprof:oso/9780199213900.001.0001},
    url = {https://doi.org/10.1093/acprof:oso/9780199213900.001.0001},
}

@article{Manzano_2020,
   title={A short introduction to the Lindblad master equation},
   volume={10},
   ISSN={2158-3226},
   url={http://dx.doi.org/10.1063/1.5115323},
   DOI={10.1063/1.5115323},
   number={2},
   journal={AIP Advances},
   publisher={AIP Publishing},
   author={Manzano, Daniel},
   year={2020},
   month=Feb }

@book{10.1093/acprof:oso/9780199212521.001.0001,
    author = {Kiefer, Claus},
    title = {Quantum Gravity},
    publisher = {Oxford University Press},
    year = {2007},
    month = {02},
    isbn = {9780199212521},
    doi = { 10.1093/acprof:oso/9780199212521.001.0001},
    url = {               https://doi.org/10.1093/acprof:oso/9780199212521.001.0001},
}

@article{Taveras_2008,
   title={Corrections to the Friedmann equations from loop quantum gravity for a universe with a free scalar field},
   volume={78},
   ISSN={1550-2368},
   url={http://dx.doi.org/10.1103/PhysRevD.78.064072},
   DOI={10.1103/physrevd.78.064072},
   number={6},
   journal={Physical Review D},
   publisher={American Physical Society (APS)},
   author={Taveras, Victor},
   year={2008},
   month=Sept }

@misc{zurek2003decoherencetransitionquantumclassical,
      title={Decoherence and the transition from quantum to classical -- REVISITED}, 
      author={Wojciech H. Zurek},
      year={2003},
      eprint={quant-ph/0306072},
      archivePrefix={arXiv},
      primaryClass={quant-ph},
      url={https://arxiv.org/abs/quant-ph/0306072}, 
}

@article{Zurek_2022,
   title={Quantum Theory of the Classical: Einselection, Envariance, Quantum Darwinism and Extantons},
   volume={24},
   ISSN={1099-4300},
   url={http://dx.doi.org/10.3390/e24111520},
   DOI={10.3390/e24111520},
   number={11},
   journal={Entropy},
   publisher={MDPI AG},
   author={Zurek, Wojciech Hubert},
   year={2022},
   month=Oct, pages={1520} }

@article{Blencowe_2013,
   title={Effective Field Theory Approach to Gravitationally Induced Decoherence},
   volume={111},
   ISSN={1079-7114},
   url={http://dx.doi.org/10.1103/PhysRevLett.111.021302},
   DOI={10.1103/physrevlett.111.021302},
   number={2},
   journal={Physical Review Letters},
   publisher={American Physical Society (APS)},
   author={Blencowe, M. P.},
   year={2013},
   month=July }

@article{Domi_2024,
   title={Understanding gravitationally induced decoherence parameters in neutrino oscillations using a microscopic quantum mechanical model},
   volume={2024},
   ISSN={1475-7516},
   url={http://dx.doi.org/10.1088/1475-7516/2024/11/006},
   DOI={10.1088/1475-7516/2024/11/006},
   number={11},
   journal={Journal of Cosmology and Astroparticle Physics},
   publisher={IOP Publishing},
   author={Domi, Alba and Eberl, Thomas and Fahn, Max Joseph and Giesel, Kristina and Hennig, Lukas and Katz, Ulrich and Kemper, Roman and Kobler, Michael},
   year={2024},
   month=Nov, pages={006} }

@misc{fahn2026generalisinggravitationallyinduceddecoherence,
      title={Generalising gravitationally induced decoherence beyond linear environmental interactions in a microscopic quantum mechanical toy model}, 
      author={Max Joseph Fahn and Renata Ferrero and Kristina Giesel and Roman Kemper},
      year={2026},
      eprint={2605.25936},
      archivePrefix={arXiv},
      primaryClass={gr-qc},
      url={https://arxiv.org/abs/2605.25936}, 
}

@misc{li2019preinflationaryperturbationsthermalbackground,
      title={Preinflationary Perturbations in The Thermal Background}, 
      author={Zucheng Li and Yanbin Deng and Yong-chang Huang},
      year={2019},
      eprint={1903.00930},
      archivePrefix={arXiv},
      primaryClass={gr-qc},
      url={https://arxiv.org/abs/1903.00930}, 
}

@article{Bhattacharya_2006,
   title={Temperature of the Inflaton and Duration of Inflation from Wilkinson Microwave Anisotropy Probe Data},
   volume={96},
   ISSN={1079-7114},
   url={http://dx.doi.org/10.1103/PhysRevLett.96.121302},
   DOI={10.1103/physrevlett.96.121302},
   number={12},
   journal={Physical Review Letters},
   publisher={American Physical Society (APS)},
   author={Bhattacharya, Kaushik and Mohanty, Subhendra and Rangarajan, Raghavan},
   year={2006},
   month=Mar }

@article{Wang_2017,
   title={Thermal gravitational-wave background in the general pre-inflationary scenario},
   volume={2017},
   ISSN={1475-7516},
   url={http://dx.doi.org/10.1088/1475-7516/2017/01/053},
   DOI={10.1088/1475-7516/2017/01/053},
   number={01},
   journal={Journal of Cosmology and Astroparticle Physics},
   publisher={IOP Publishing},
   author={Wang, Kai and Santos, Larissa and Xia, Jun-Qing and Zhao, Wen},
   year={2017},
   month=Jan, pages={053–053} }

@article{Mukhanov_1997,
   title={Backreaction Problem for Cosmological Perturbations},
   volume={78},
   ISSN={1079-7114},
   url={http://dx.doi.org/10.1103/PhysRevLett.78.1624},
   DOI={10.1103/physrevlett.78.1624},
   number={9},
   journal={Physical Review Letters},
   publisher={American Physical Society (APS)},
   author={Mukhanov, V. F. and Abramo, L. Raul W. and Brandenberger, Robert H.},
   year={1997},
   month=mar, pages={1624–1627} }

@phdthesis{li2008cosmological,
  author  = {Li, Nan},
  title   = {Cosmological Backreaction: From the Local Hubble Expansion Rate to Dark Energy},
  school  = {Universit{\"a}t Bielefeld},
  address = {Bielefeld},
  year    = {2008},
  url     = {https://webdoc.sub.gwdg.de/ebook/dissts/Bielefeld/Li2008.pdf}
}

@article{Brown_1995,
   title={Dust as a standard of space and time in canonical quantum gravity},
   volume={51},
   ISSN={0556-2821},
   url={http://dx.doi.org/10.1103/PhysRevD.51.5600},
   DOI={10.1103/physrevd.51.5600},
   number={10},
   journal={Physical Review D},
   publisher={American Physical Society (APS)},
   author={Brown, J. David and Kuchař, Karel V.},
   year={1995},
   month=May, pages={5600–5629} }

@article{Husain_2012,
   title={Time and a Physical Hamiltonian for Quantum Gravity},
   volume={108},
   ISSN={1079-7114},
   url={http://dx.doi.org/10.1103/PhysRevLett.108.141301},
   DOI={10.1103/physrevlett.108.141301},
   number={14},
   journal={Physical Review Letters},
   publisher={American Physical Society (APS)},
   author={Husain, Viqar and Pawłowski, Tomasz},
   year={2012},
   month=Apr }

@misc{anderson2012problemtimequantumgravity,
      title={The Problem of Time in Quantum Gravity}, 
      author={Edward Anderson},
      year={2012},
      eprint={1009.2157},
      archivePrefix={arXiv},
      primaryClass={gr-qc},
      url={https://arxiv.org/abs/1009.2157}, 
}

@article{Ali_2018,
   title={Universe as an oscillator},
   volume={98},
   ISSN={2470-0029},
   url={http://dx.doi.org/10.1103/PhysRevD.98.086002},
   DOI={10.1103/physrevd.98.086002},
   number={8},
   journal={Physical Review D},
   publisher={American Physical Society (APS)},
   author={Ali, Masooma and Hassan, Syed Moeez and Husain, Viqar},
   year={2018},
   month=Oct }

@article{Abramo_1997,
   title={Energy-momentum tensor for cosmological perturbations},
   volume={56},
   ISSN={1089-4918},
   url={http://dx.doi.org/10.1103/PhysRevD.56.3248},
   DOI={10.1103/physrevd.56.3248},
   number={6},
   journal={Physical Review D},
   publisher={American Physical Society (APS)},
   author={Abramo, L. Raul W. and Brandenberger, Robert H. and Mukhanov, V. F.},
   year={1997},
   month=Sept, pages={3248–3257} }

@misc{brandenberger2002reactioncosmologicalperturbationscosmological,
      title={Back Reaction of Cosmological Perturbations and the Cosmological Constant Problem}, 
      author={Robert H. Brandenberger},
      year={2002},
      eprint={hep-th/0210165},
      archivePrefix={arXiv},
      primaryClass={hep-th},
      url={https://arxiv.org/abs/hep-th/0210165}, 
}

@misc{baumann2012tasilecturesinflation,
      title={TASI Lectures on Inflation}, 
      author={Daniel Baumann},
      year={2012},
      eprint={0907.5424},
      archivePrefix={arXiv},
      primaryClass={hep-th},
      url={https://arxiv.org/abs/0907.5424}, 
}

@article{Geshnizjani_2005,
   title={Back-reaction of perturbations in two scalar field inflationary models},
   volume={2005},
   ISSN={1475-7516},
   url={http://dx.doi.org/10.1088/1475-7516/2005/04/006},
   DOI={10.1088/1475-7516/2005/04/006},
   number={04},
   journal={Journal of Cosmology and Astroparticle Physics},
   publisher={IOP Publishing},
   author={Geshnizjani, Ghazal and Brandenberger, Robert},
   year={2005},
   month=Apr, pages={006–006} }

@article{ refId0,
	author = {{Planck Collaboration} and {Akrami, Y.} and {Arroja, F.} and {Ashdown, M.} and {Aumont, J.} and {Baccigalupi, C.} and {Ballardini, M.} and {Banday, A. J.} and {Barreiro, R. B.} and {Bartolo, N.} and {Basak, S.} and {Benabed, K.} and {Bernard, J.-P.} and {Bersanelli, M.} and {Bielewicz, P.} and {Bock, J. J.} and {Bond, J. R.} and {Borrill, J.} and {Bouchet, F. R.} and {Boulanger, F.} and {Bucher, M.} and {Burigana, C.} and {Butler, R. C.} and {Calabrese, E.} and {Cardoso, J.-F.} and {Carron, J.} and {Challinor, A.} and {Chiang, H. C.} and {Colombo, L. P. L.} and {Combet, C.} and {Contreras, D.} and {Crill, B. P.} and {Cuttaia, F.} and {de Bernardis, P.} and {de Zotti, G.} and {Delabrouille, J.} and {Delouis, J.-M.} and {Di Valentino, E.} and {Diego, J. M.} and {Donzelli, S.} and {Doré, O.} and {Douspis, M.} and {Ducout, A.} and {Dupac, X.} and {Dusini, S.} and {Efstathiou, G.} and {Elsner, F.} and {Enßlin, T. A.} and {Eriksen, H. K.} and {Fantaye, Y.} and {Fergusson, J.} and {Fernandez-Cobos, R.} and {Finelli, F.} and {Forastieri, F.} and {Frailis, M.} and {Franceschi, E.} and {Frolov, A.} and {Galeotta, S.} and {Galli, S.} and {Ganga, K.} and {Gauthier, C.} and {Génova-Santos, R. T.} and {Gerbino, M.} and {Ghosh, T.} and {González-Nuevo, J.} and {Górski, K. M.} and {Gratton, S.} and {Gruppuso, A.} and {Gudmundsson, J. E.} and {Hamann, J.} and {Handley, W.} and {Hansen, F. K.} and {Herranz, D.} and {Hivon, E.} and {Hooper, D. C.} and {Huang, Z.} and {Jaffe, A. H.} and {Jones, W. C.} and {Keihänen, E.} and {Keskitalo, R.} and {Kiiveri, K.} and {Kim, J.} and {Kisner, T. S.} and {Krachmalnicoff, N.} and {Kunz, M.} and {Kurki-Suonio, H.} and {Lagache, G.} and {Lamarre, J.-M.} and {Lasenby, A.} and {Lattanzi, M.} and {Lawrence, C. R.} and {Le Jeune, M.} and {Lesgourgues, J.} and {Levrier, F.} and {Lewis, A.} and {Liguori, M.} and {Lilje, P. B.} and {Lindholm, V.} and {López-Caniego, M.} and {Lubin, P. M.} and {Ma, Y.-Z.} and {Macías-Pérez, J. F.} and {Maggio, G.} and {Maino, D.} and {Mandolesi, N.} and {Mangilli, A.} and {Marcos-Caballero, A.} and {Maris, M.} and {Martin, P. G.} and {Martínez-González, E.} and {Matarrese, S.} and {Mauri, N.} and {McEwen, J. D.} and {Meerburg, P. D.} and {Meinhold, P. R.} and {Melchiorri, A.} and {Mennella, A.} and {Migliaccio, M.} and {Mitra, S.} and {Miville-Deschênes, M.-A.} and {Molinari, D.} and {Moneti, A.} and {Montier, L.} and {Morgante, G.} and {Moss, A.} and {Münchmeyer, M.} and {Natoli, P.} and {Nørgaard-Nielsen, H. U.} and {Pagano, L.} and {Paoletti, D.} and {Partridge, B.} and {Patanchon, G.} and {Peiris, H. V.} and {Perrotta, F.} and {Pettorino, V.} and {Piacentini, F.} and {Polastri, L.} and {Polenta, G.} and {Puget, J.-L.} and {Rachen, J. P.} and {Reinecke, M.} and {Remazeilles, M.} and {Renzi, A.} and {Rocha, G.} and {Rosset, C.} and {Roudier, G.} and {Rubiño-Martín, J. A.} and {Ruiz-Granados, B.} and {Salvati, L.} and {Sandri, M.} and {Savelainen, M.} and {Scott, D.} and {Shellard, E. P. S.} and {Shiraishi, M.} and {Sirignano, C.} and {Sirri, G.} and {Spencer, L. D.} and {Sunyaev, R.} and {Suur-Uski, A.-S.} and {Tauber, J. A.} and {Tavagnacco, D.} and {Tenti, M.} and {Toffolatti, L.} and {Tomasi, M.} and {Trombetti, T.} and {Valiviita, J.} and {Van Tent, B.} and {Vielva, P.} and {Villa, F.} and {Vittorio, N.} and {Wandelt, B. D.} and {Wehus, I. K.} and {White, S. D. M.} and {Zacchei, A.} and {Zibin, J. P.} and {Zonca, A.}},
	title = {Planck 2018 results - X. Constraints on inflation},
	DOI= "10.1051/0004-6361/201833887",
	url= "https://doi.org/10.1051/0004-6361/201833887",
	journal = {A\&A},
	year = 2020,
	volume = 641,
	pages = "A10",
}

@article{CALDEIRA1983587,
title = {Path integral approach to quantum Brownian motion},
journal = {Physica A: Statistical Mechanics and its Applications},
volume = {121},
number = {3},
pages = {587-616},
year = {1983},
issn = {0378-4371},
doi = {10.1016/0378-4371(83)90013-4},
url = {https://www.sciencedirect.com/science/article/pii/0378437183900134},
author = {A.O. Caldeira and A.J. Leggett}
}

@misc{unruh1998cosmologicallongwavelengthperturbations,
      title={Cosmological long wavelength perturbations}, 
      author={W. Unruh},
      year={1998},
      eprint={astro-ph/9802323},
      archivePrefix={arXiv},
      primaryClass={astro-ph},
      url={https://arxiv.org/abs/astro-ph/9802323}, 
}

@article{Giesel_2020,
   title={Mukhanov-Sasaki equation in a manifestly gauge-invariant linearized cosmological perturbation theory with dust reference fields},
   volume={102},
   ISSN={2470-0029},
   url={http://dx.doi.org/10.1103/PhysRevD.102.023524},
   DOI={10.1103/physrevd.102.023524},
   number={2},
   journal={Physical Review D},
   publisher={American Physical Society (APS)},
   author={Giesel, Kristina and Herold, Laura and Li, Bao-Fei and Singh, Parampreet},
   year={2020},
   month=July }

@article{Schlosshauer_2019,
   title={Quantum decoherence},
   volume={831},
   ISSN={0370-1573},
   url={http://dx.doi.org/10.1016/j.physrep.2019.10.001},
   DOI={10.1016/j.physrep.2019.10.001},
   journal={Physics Reports},
   publisher={Elsevier BV},
   author={Schlosshauer, Maximilian},
   year={2019},
   month=Oct, pages={1–57} }

@article{Kiefer_1987,
   doi = {10.1088/0264-9381/4/5/031},
   url = {https://doi.org/10.1088/0264-9381/4/5/031},
   year = {1987},
   month = {sep},
   publisher = {},
   volume = {4},
   number = {5},
   pages = {1369},
   author = {C Kiefer},
   title = {Continuous measurement of mini-superspace variables by higher multipoles},
   journal = {Classical and Quantum Gravity}
}

@article{burgess2014efthorizonstochasticinflation,
  author  = {Burgess, C. P. and Holman, R. and Tasinato, G. and Williams, M.},
  title   = {EFT beyond the horizon: stochastic inflation and how primordial quantum fluctuations go classical},
  journal = {Journal of High Energy Physics},
  year    = {2015},
  volume  = {2015},
  number  = {3},
  pages   = {90},
  doi     = {10.1007/JHEP03(2015)090},
}

@article{Martin_2018,
   title={Observational constraints on quantum decoherence during inflation},
   volume={2018},
   ISSN={1475-7516},
   url={http://dx.doi.org/10.1088/1475-7516/2018/05/063},
   DOI={10.1088/1475-7516/2018/05/063},
   number={05},
   journal={Journal of Cosmology and Astroparticle Physics},
   publisher={IOP Publishing},
   author={Martin, Jérôme and Vennin, Vincent},
   year={2018},
   month=May, pages={063–063} }

@article{Colas_2022,
   title={Benchmarking the cosmological master equations},
   volume={82},
   ISSN={1434-6052},
   url={http://dx.doi.org/10.1140/epjc/s10052-022-11047-9},
   DOI={10.1140/epjc/s10052-022-11047-9},
   number={12},
   journal={The European Physical Journal C},
   publisher={Springer Science and Business Media LLC},
   author={Colas, Thomas and Grain, Julien and Vennin, Vincent},
   year={2022},
   month=Dec }

@article{Burgess_2023,
   title={Minimal decoherence from inflation},
   volume={2023},
   ISSN={1475-7516},
   url={http://dx.doi.org/10.1088/1475-7516/2023/07/022},
   DOI={10.1088/1475-7516/2023/07/022},
   number={07},
   journal={Journal of Cosmology and Astroparticle Physics},
   publisher={IOP Publishing},
   author={Burgess, C.P. and Holman, R. and Kaplanek, Greg and Martin, Jérôme and Vennin, Vincent},
   year={2023},
   month=July, pages={022} }

@article{Geshnizjani_2002,
   title={Back reaction and the local cosmological expansion rate},
   volume={66},
   ISSN={1089-4918},
   url={http://dx.doi.org/10.1103/PhysRevD.66.123507},
   DOI={10.1103/physrevd.66.123507},
   number={12},
   journal={Physical Review D},
   publisher={American Physical Society (APS)},
   author={Geshnizjani, Ghazal and Brandenberger, Robert},
   year={2002},
   month=Dec }

@article{PhysRevLett.70.1187,
  title = {Coherent states via decoherence},
  author = {Zurek, Wojciech H. and Habib, Salman and Paz, Juan Pablo},
  journal = {Phys. Rev. Lett.},
  volume = {70},
  issue = {9},
  pages = {1187--1190},
  numpages = {0},
  year = {1993},
  month = {Mar},
  publisher = {American Physical Society},
  doi = {10.1103/PhysRevLett.70.1187},
  url = {https://link.aps.org/doi/10.1103/PhysRevLett.70.1187}
}

@mastersthesis{Kulyabin:2025MSc,
  author  = {Alexander Kulyabin},
  title   = {Decoherence Model in Quantum Cosmology},
  school  = {Friedrich-Alexander-Universit{\"a}t Erlangen-N{\"u}rnberg},
  year    = {2025},
  type    = {Master's thesis}
}

@article{Corichi_2011,
   title={Coherent semiclassical states for loop quantum cosmology},
   volume={84},
   ISSN={1550-2368},
   url={http://dx.doi.org/10.1103/PhysRevD.84.044021},
   DOI={10.1103/physrevd.84.044021},
   number={4},
   journal={Physical Review D},
   publisher={American Physical Society (APS)},
   author={Corichi, Alejandro and Montoya, Edison},
   year={2011},
   month=Aug }
\end{document}